%% file: 2012mehani_oml_performance_RT.tex
\documentclass[twoside, openright]{nictatechreport}

\usepackage{graphicx} 
\usepackage{subfigure} 
\usepackage{url}
\usepackage{amsmath} 
\usepackage{booktabs}
\usepackage{multirow}
\usepackage{moreverb}
\usepackage{rotating}
\usepackage{slashbox}
\usepackage{textcomp}
\usepackage{array}
\usepackage{color}
\usepackage[maxnames=4,style=numeric,firstinits=true,sortcites,url=false,useeditor=false]{biblatex} \bibliography{2012mehani_oml_performance}
\newcommand{\theTitle}{A Method for the Characterisation of Observer Effects
and its Application to OML}

\newcommand{\theAuthor}{Olivier Mehani, Guillaume Jourjon and Thierry Rakotoarivelo}
\newcommand{\theKeywords}{Network Measurements, OML, Performance Analysis}

\usepackage[pdftex, colorlinks=true, pdfstartview=FitV, linkcolor=black, citecolor=black, urlcolor=blue, unicode,
  pdftitle={\theTitle},
  pdfauthor={\theAuthor},
  pdfkeywords={\theKeywords}]{hyperref}
\usepackage{pdflscape}

\usepackage{aeguill}

\def\texturL#1{\texttt{\url{#1}}\endgroup}

\newcommand{\eg}{\textit{e.g.}}
\newcommand{\ie}{\textit{i.e.}}

\newcommand{\apriori}{\textit{a priori}}

\newcommand{\treatment}[1]{\textsf{#1}}
\newcommand\nooml{\treatment{nooml}}
\newcommand\oml{\treatment{oml}}
\newcommand\ooml{\treatment{o}}
\newcommand\Ooml{\treatment{O}}
\newcommand\Ofoml{\treatment{Of}}
\newcommand\nothreads{\treatment{nothreads}}
\newcommand\threads{\treatment{threads}}
\newcommand\ftn{\treatment{10}}
\newcommand\fft{\treatment{50}}
\newcommand\fsf{\treatment{75}}
\newcommand\fh{\treatment{100}}
\newcommand\fth{\treatment{200}}
\newcommand\ftth{\treatment{300}}

\newcommand{\syscmd}[1]{\texttt{#1}}
\newcommand\tcpdump{\syscmd{tcpdump}}
\newcommand\noomltrace{\syscmd{trace-nooml}}
\newcommand\omltrace{\syscmd{trace-oml2}}

\title{\theTitle}
\author{Olivier Mehani$^{\star,1}$, Guillaume Jourjon$^1$, Thierry Rakotoarivelo$^1$}

\otheraffiliation{$^\star$Corresponding author: olivier.mehani@nicta.com.au\\ 
$^1$first.last@nicta.com.au }

\submitdate{MSWIM 2012, May 13, 2012}
\reportnumber{TR-5895}
\pubhistory{Submitted to MSWIM 2012}

\begin{document}

\frontmatter

\pagestyle{empty}


\input{2012mehani_oml_performance_abstract}
\input{2012mehani_oml_performance_introduction}

\input{2012mehani_oml_performance_method}

\input{2012mehani_oml_performance_results}

\input{2012mehani_oml_performance_discussion}

\input{2012mehani_oml_performance_related}
\input{2012mehani_oml_performance_conclusion}

\printbibliography

\end{document}

%% file: 2012mehani_oml_performance_abstract.tex
\begin{abstract}

In all measurement campaigns, one needs to assert that the instrumentation tools
do not significantly impact the system being monitored. This is critical to
future claims based on the collected data and is sometimes overseen in
experimental studies.  We propose a method to evaluate the potential ``observer
effect'' of an instrumentation system, and apply it to the OMF Measurement
Library (OML).  OML allows the instrumentation of almost any software to collect
any type of measurements.  As it is increasingly being used in networking
research, it is important to characterise possible biases it may
introduce in the collected metrics.  Thus, we study its effect on multiple types
of reports from various applications commonly used in wireless research. To this
end, we designed experiments comparing OML-instrumented software with their
original flavours.
%
%
%
%
Our analyses of the results from these experiments show that, with an appropriate
reporting setup, OML has no significant impact on the instrumented applications, and may even improve some of their performances in specifics cases.
%
We discuss our methodology and the implication of using OML, and provide
guidelines on instrumenting off-the-shelf software.


\end{abstract}

%% file: 2012mehani_oml_performance_introduction.tex
\section{Introduction}

Measurement is a foundation stone of scientific research. It is the analysis of
measured data that allows researchers to support or refute scientific claims. 
Many types of data can be collected using tools with various characteristics
in accuracy, precision, and impact on the systems under study. Without 
characterising these measurement tools, it is impossible to assess the validity 
of the conclusions that are based on the collected data.

Networking technologies are closely linked to the software system that embodies
and controls them. Many of the tools used to observe networking behaviours
consist in software executing alongside the systems under study, on the same
machine and operating environment. If not designed carefully, such tools may
alter the system's performance or functionality, which in turn may impact the
observations being made. It is therefore important to characterise the
``observer effect'' that an instrumentation software tool has on a networking 
system. This step is sometimes overseen in experimental studies.

When designing an experiment, a researcher often has to collect measurements
from multiple sources involved in the study.  In wireless
networking, this usually involves a combination of various third-party and 
home-brew software tools to measure and report the data.
OML\footnote{\url{http://oml.mytestbed.net}} is an open source measurement 
framework, which
facilitates these two phases by allowing an experimenter to collect any type of
measurements from many types of software, and store them in a unified format for
future analysis~\cite{2010white_oml}. It is composed of a client library and a
collection server. The client library is used to instrument any
software from which a researcher would like to collect measurements. It can
apply some processing (\ie, filter) to the samples on the fly, then streams them
to a collection server, which stores them in a database along with measurements
from other elements involved in the experiment. An increasing
number of researchers from various institutions have used OML in their
wireless studies~\cite{2010mathur_parknet,2011boulis_wsn_bridge}.

Our objective in this paper is to characterise the observer effect that OML
has on the software that it instruments. To this end, we selected two
types of widely used software, a network probing tool
(Iperf~\cite{2004gates_iperf2}) and a
packet capture library (libtrace~\cite{2012alcock_libtrace}). We instrumented
them using OML and compared many indicator variables between the instrumented
and vanilla flavours to detect any significant impact that the
OML instrumentation could have on the software.

Our contribution is twofold. First, we propose a methodology to characterise the
observer effect of a networking measurement framework. This methodology and its
specific application, in this paper, to OML are described
in Section~\ref{sec:method}. Second, we identify the cases where OML 
introduces statistically significant deviations in the performance of
instrumented applications, and those where no impact can be detected. This demonstration is based on the analysis of the
indicator variables selected for each instrumented tool, and is presented
in Section~\ref{sec:results}. We believe that this contribution is novel and
relevant as few other works thoroughly study the observer effect of measurement 
frameworks, and OML's use is growing within the wireless research community.
We further discuss the findings and limitations of our
study in Section~\ref{sec:discussion}, then present some related work in
Section~\ref{sec:related}, and finally conclude this paper in
Section~\ref{sec:conclusion}.

%% file: 2012mehani_oml_performance_method.tex
\section{Methodology}
\label{sec:method}

\subsection{Objectives}
\label{sec:method:objective}

Our objective is to test whether using OML version 2.6.1 to instrument a piece
of software for a networking study has any impact on its behaviours and
observed performance. If there were, this could alter the variables being
measured in the study. In other words, we would like to demonstrate that the
``observer effect'' of OML is negligible. What we define as an impact is a
statistically significant deviation in some performance indicators between the
original software and other OML-instrumented flavours.

While several researchers have used OML to instrument many software tools, we
limit this study to two of them, namely the Iperf network probing
application~\cite{2004gates_iperf2} , and the libtrace~\cite{2012alcock_libtrace} 
packet
capture library. It is not the purpose of our study to evaluate the accuracy or
precision of these tools. Rather, we are only interested in possible deviations
introduced by the use of OML instead of their original reporting channels.

We propose the following four-step method to characterise the observer 
effect of OML.
\begin{enumerate}

  \item formulation of the objective and selection of the material to use;

  \item design of the experiment, which includes the formulation of the
    hypotheses, the identification of the factors which may influence these
    hypotheses and the dependent variables to measure in order to test them; the
    description of the experiment setup is also considered there;

  \item the analysis of the measured variables, using statistical tools which
    are adequate to the nature of the selected factors and variables;

  \item the discussion on the conclusions from the results of step 3, and the
    limitations of step 2, which may lead to further study via another iteration
    of step 1.

\end{enumerate}

\subsection{Materials}

\subsubsection{OML}

OML~\cite{2010white_oml} is a distributed, open source, and multithreaded
measurement framework. It provides a client library, which can be used to
instrument any piece of software. An instrumented application ``injects'' its
samples to the library for processing and streaming to at least one collection
server. When requested by the software's user, this client library may apply
some processing filters to the measurements before forwarding them.  The
collection server receives the samples from software running on all nodes involved
in a given experiment (identified by a unique ID), and store them in a
timestamped database
with a unified format.

OML has an active user community, which has contributed to the instrumentation
of a wide range of software used in networking research,\footnote{\url{http://omlapp.mytestbed.net}} such as the radiotap library (802.11 frame characteristics), wlanconfig (driver status), but also applications such as VLC (media streaming) or btclient (BitTorrent
client).

Within an experiment, each OML-instrumented software generates a timestamp
(\verb#oml_ts_client#) for each generated sample. The receiving server rebases
this timestamp to a local origin (\verb#oml_ts_server#), \ie, the time of the
first connected client for that experiment. This allows for low-resolution
($\sim$1\,s) comparisons and correlations between measurements from different
sources. Since the client and rebased timestamps are both stored in the
database, a high-resolution comparison can also be achieved by deploying a
separate synchronisation scheme such as~\cite{2009veitch_robust_synchronisation} on all clients. 

To instrument an application with OML, a developer first defines one or many
measurement points (MPs) within its source code. An MP is an abstraction for a
tuple of related metrics to be reported at the same instant. Thus the MPs define
all the potential measurements that the application can report. The modified
source code is then compiled against the OML client library to generate the
OML-instrumented software.  At run-time, the experimenter can request some or
all of the defined MPs to generate measurement streams (MSs). This is done
through an XML configuration file passed to the software at startup. 

\begin{figure}[tb!]
 \centering
 \resizebox{\columnwidth}{!}{\input{oml-datapath.pdftex_t}}

 \caption{Measurement data path in OML: Three MPs are filtered to generate MSs
 which get sent for storage at different locations.}

 \label{fig:oml-datapath}
\end{figure}
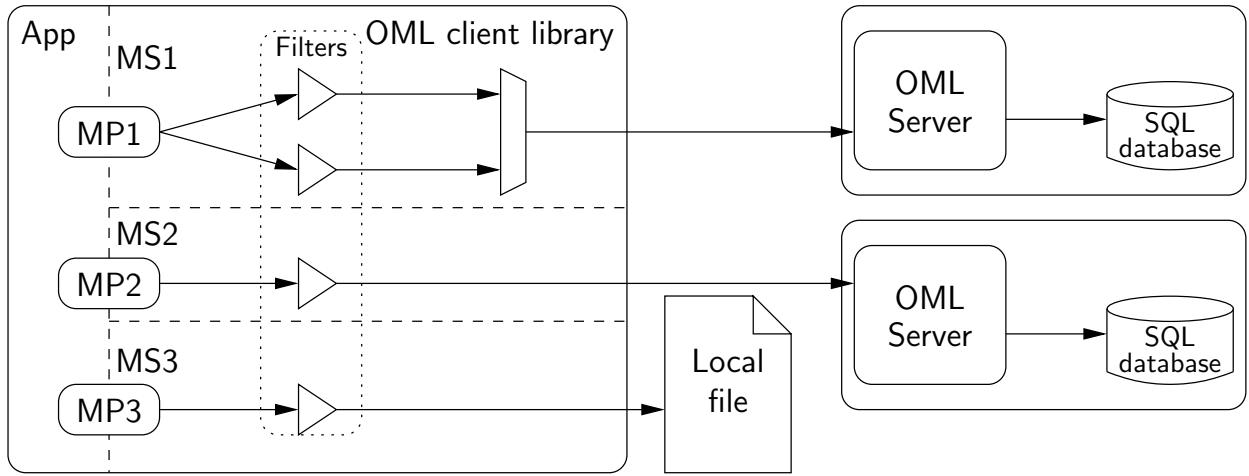

Prior to streaming MSs to a server, the client library may apply predefined 
filters to the samples, as described in the XML configuration file. This filtering 
process is illustrated in~\figurename~\ref{fig:oml-datapath}. Filters are 
composable functions, which are applied to a subset of metrics from MSs over a 
given time or sample period (\eg, every 1\,s or 10 samples). Thus they integrate 
incoming MSs into newly generated outgoing MSs. OML has some built-in filters 
and allow users to create custom ones. Examples of simple OML filtering
capabilities include
averaging a metric over a time window, or getting its extreme values (min/max) over 
a given number of samples.

OML clients stream MSs to the servers over TCP using a custom protocol.
They have a finite-sized outgoing buffers and may therefore drop some
measurement samples if the path to the server cannot provide sufficient capacity
to cater for the sample rate. A sequence ID incremented for every sample allows
the detection of such events on the server side.

\subsubsection{Network Probing with Iperf}
\label{sec:iperf}

Iperf~\cite{2004gates_iperf2} is a versatile open source active network probing tool. It allows an
experimenter to test the characteristics of a network path using either TCP or
UDP. Its code is multithreaded to limit the impact of reporting---either on the
console or in a CSV file---on the high-speed generation of probe packets.

Iperf can report a number of metrics depending on the transport protocol in use.
For TCP only the transferred size, from which the throughput is derived, can be
observed. 
For UDP%
, packet loss and jitter information can also be reported.  
The periodicity of Iperf's reports is
configurable from once for an entire run to every half a second. The internal
aggregation function depends on the metric: the transferred size and losses are
summed, while the latest value to date is reported for the jitter.

There tends to be some confusion with the definition of jitter~\cite{rfc3393}.
In the case of Iperf, the term refers to the variation in packets transit times,
as proposed in~\cite{rfc1889}, and it is computed at packet $p$ as
\begin{gather}
  \tau_p = t_p^{Rcv} - t_p^{Snd} \notag\\
  \Delta\tau_p = |\tau_p - \tau_{p-1}| \notag\\
  J_p = J_{p-1} + \frac1{16}(\Delta\tau_p - J_{p-1}).
  \label{eq:jitter}
\end{gather}
As this jitter itself is based on the \emph{variation} of transit times, rather
than the immediate values, it is rather robust to loose time synchronisation
between sender and receiver. 


We have instrumented version 2.0.5 of Iperf to support reporting via
OML.\footnote{\url{http://omlapp.mytestbed.net/projects/iperf/wiki}}
We implemented two new additional forms of measurement
reporting styles, \emph{legacy} (\syscmd{iperf -y o}) and \emph{advanced}
(\syscmd{iperf -y O}), which differ in the amount of processing that is done in
the application. Table~\ref{tab:iperf-mp} summarises the performance metrics
directly reported by the different flavours of Iperf used in this study.

\begin{table}
  \addtolength{\tabcolsep}{-.35em}
  \centering
  \caption{Metrics reported by the various Iperf flavours in this study}
  \label{tab:iperf-mp}
  \begin{tabular}{ll}
    \toprule
    \multicolumn{1}{c}{\textbf{Flavour}} & \multicolumn{1}{c}{\textbf{Reported
    metrics (for UDP traffic)}} \\
    \midrule
    Vanilla	& Transferred size,	throughput,	losses, jitter \\ 
    OML legacy	& Transferred size,	throughput,	losses, jitter \\ 
    OML advanced	& Packet ID, size, emission and reception\\ 
    & timestamps\\
    \bottomrule
  \end{tabular}
\end{table}

In the legacy mode, the aggregation of the measurements is done using Iperf's
standard code, and the periodic reports are sent out through OML via three MPs:
\verb_transfer_ for the size, \verb_losses_ for lost and sent datagrams, and
\verb_jitter_ for Iperf's implementation of~\eqref{eq:jitter}. In the advanced
mode, Iperf directly reports information about each packet sent or received via
OML, in the \verb_packets_ MP which contain identification, size and both sent
and received (if relevant) times for each packet, down to the microsecond. The
advanced mode is more in line with OML's approach, where the measured data is
reported verbatim by the application and all processing and consolidation is
done through filters, thus allowing more experiment-specific treatment without
impacting the main operation of the application. Noting that, in most of the
literature using Iperf, there is a lack of precise reporting of versions,
platforms and parameters, we also implemented MPs to report such ancillary
information about the experiment as the version numbers and command line
arguments.


\subsubsection{Packet Capture with libtrace}
\label{sec:tool-capture}

Packet capture in networking environments is usually done via wrapper libraries
hiding the operating system's underlying API. Perhaps the most common libraries
for this purpose are libpcap\footnote{\url{http://tcpdump.org}} and libtrace
\cite{2012alcock_libtrace}. The latter offers a broader range of input and 
output APIs and formats than the former.


We implemented
\omltrace,\footnote{\url{http://omlapp.mytestbed.net/projects/omlapp/repository/revisions/master/show/trace}} a simple packet-capturing application which uses
libtrace to get packet header information from the kernel.  This data is then
available from \omltrace's MPs, and can be collected using OML as required by
the experimenter.  Table~\ref{tab:trace-mp} lists the available metrics from the
packet headers, and \omltrace's MPs providing them. As previously, we need to
consider different flavours of this application, and also implemented an almost
identical \noomltrace, which reports the metrics in a local CSV file rather than
through OML, by modifying the reporting functions.

 


\begin{table}[tb!]
  \centering
  \addtolength{\tabcolsep}{-.5em}
  \caption{Measurement Points \& Metrics available from \omltrace.}
  \label{tab:trace-mp}
  \begin{tabular}{ll} \\
    \toprule
    \multicolumn{1}{c}{\textbf{MP name}} & \multicolumn{1}{c}{\textbf{Reported metrics}} \\
    \midrule
   	\verb_radiotap_	& Seq. ID, MAC addresses, \& other MAC fields\\
   \cmidrule{1-2}  
   \verb_ip_ 	& Pkt ID, IP addresses, length, \& other IP fields\\
   \cmidrule{1-2}
   \verb_tcp_ or \verb_udp_ & ID, length, ports, and other transport fields\\
   \cmidrule{1-2} 
    All MPs & Timestamp	\\
    \bottomrule
  \end{tabular}
\end{table}

\subsection{Experimental Design}
\label{sec:experiments}

To achieve our goal (section \ref{sec:method:objective}), we compare some
performance indicators between the original and OML instrumented flavours of
Iperf and our libtrace-based tool. 
In addition, we also quantify the impact of the forwarding of MSs on 
other experiment-generated traffic in a shared wireless environment. 
Thus, we design and perform three sets of
experiments, one focusing on Iperf, another on libtrace, and the last one on the
impact of reporting traffic. 

The two first sets are based on a simple 2-hop topology composed of a sender
($Snd$), a router ($Rtr$; with an ingress $RtrIn$ and egress $RtrEg$ interfaces),
and a receiver ($Rcv$). Within each set of experiments, we vary some specific
setup parameters, such as the flavour of the tool under study (\eg, vanilla or
OML-instrumented with feature $X$ enabled). These parameters are our study's
independent variables (or factors). For each experiment trial, we measure some
performance indicators such as the sending rate or report accuracy, which are
our study's dependent variables. In these two sets, the measurement streams are
sent to an OML server on a separate network from the experiment-generated
traffic, and thus do not impact it.

The third set involves a sender ($Snd$) generating traffic towards a a receiver 
($Rcv$). The same OML-instrumented tools as previously are used to collect 
measurements. However, in this set the measurement streams are 
sent to an OML server over the same wireless channel as the generated traffic 
between $Snd$ and $Rcv$.

\subsubsection{Experiment Set 1}

This set aims at characterising the effect of OML on the performance of Iperf.
We are specifically interested in how Iperf's traffic generation and the accuracy
of its traffic statistic reports may be altered by the OML instrumentation. Thus
our working null hypothesis for this first set is: ``the OML instrumentation of
Iperf has no significant effect on its packet-sending rate nor on the accuracy
of its throughput and jitter reports.''

\paragraph{Experimental Factors} 

Many factors may influence the performance or accuracy of Iperf. For this first
set, we restrict our experimental design to three factors.  The first factor is
the \textbf{Iperf flavour} being used. Indeed, the use of OML and its various
features introduce some processing overheads on Iperf, which may deteriorate
performance and report accuracy, as compared to the original Iperf. We consider
the following values for this factor: 
\begin{description}
  \item[\nooml] the original Iperf, reporting to a CSV file
  \item[\ooml] OML-enabled Iperf, reporting metrics computed by Iperf (the same as the 
    original version)
  \item[\Ooml] OML-enabled Iperf, reporting information on all sent/received packets
  \item[\Ofoml] OML-enabled Iperf, reporting metrics computed by OML filters
\end{description}

The second factor we are interested in is the \textbf{set rate} at which Iperf
is instructed to generate the experimental traffic. At high rates, the per-packet instrumentation processing may take
longer than the inter-packet sending interval, thus impeding the actual sending
rate.  Although the sending rate is a continuous variable, for the purpose of
our study we are only interested in given rate values, and thus treat it as a
fixed factor with the values: \ftn, \fft, \fsf, \fh, \fth\ and \ftth\,Mbps.

The last factor is the use of \textbf{threads}. The original Iperf has the option
to use threads or not to report traffic information out of the main traffic
generation loop. In our study, this feature may mask the effects introduced by
the OML-instrumentation. Thus we consider two values, \threads\ and \nothreads.

\paragraph{Response Variables}
In this set of experiments, we decide to measure the following three dependent
variables, which we will use in the next section to test our working hypothesis.
First, we measure the \textbf{actual sending rate}  ($R_{t}^{RtrIn}$) of the
sending Iperf as computed from \tcpdump\ traces ($t$) on the router's ingress link from
the sender. 

We also measure the \textbf{accuracy of the throughput report}
($T_{Diff}^{Rcv}$) of the receiving Iperf, which is the difference between the
throughput as reported by the receiving Iperf ($i$) and the computed throughput
from \tcpdump\ traces on the router's egress link to the receiver. For a
given sample, $T_{Diff}^{Rcv} = R_{t}^{RtrEg} - T_{i}^{Rcv}$.

Finally, we
measure the \textbf{jitter report accuracy} ($J_{Diff}$) of the receiving Iperf,
which is the difference between the jitter as reported by the receiving Iperf
and the computed jitter using equation~\eqref{eq:jitter} on the \tcpdump\ traces
on the router's egress link to the receiver. For given sample, we have $J_{Diff} =
J_{i}^{Rcv} - J_{t}^{RtrEg}$.

\subsubsection{Experiment Set 2}

The goal of this set is to evaluate the effect of OML on the a tool based on the
libtrace packet-capture library. We are particularly interested in finding out
if the OML instrumentation degrades the accuracy of the packet and timestamp
reports from libtrace. Thus our working null hypothesis in this case is: ``the
OML instrumentation of libtrace has no significant effect on the accuracy of its
packet and timestamp reports.''

\paragraph{Experimental Factors}
As in the previous experiment set, the first obvious factor that may impact the
accuracy of libtrace's reports is the \textbf{librace application flavour} being
used. Here we only consider the two simple cases where libtrace is used with and
without OML instrumentation, \ie, \omltrace\ (\oml) and \noomltrace\ (\nooml).
In the former case, we also consider the use of a summation filter (\Ofoml) on the
length field of the IP header (\texttt{ip\_len}).

Similarly, the second factor that we consider is the \textbf{set rate} for the
traffic generation between the sender and the receiver. Indeed, increased packet
rate may translate in increased instrumentation processing for packet reporting
on the receiver. This may result in received packets not being reported in the
measurement file, as they are dropped by the instrumentation mechanism being
overwhelmed by the high packet rate. For this factor, we use the same fixed
rates as in the first experiment set, \ftn, \fft, \fsf, \fh, \fth\ and \ftth\,Mbps.

\paragraph{Response Variables}

We measure three dependent variables in this experiment set. The first one is the
\textbf{accuracy of packet reports}, in the form of losses ($L^{Rcv}$), generated by our
libtrace-based tool at the receiver. This is the ratio between the
number of packets sent to the receiver as counted from \tcpdump\ traces on the
router's egress link ($N_t^{RtrEg}$) to that of the packets reported by the
receiver's instrumentation ($N_i^{Rcv}$). Thus for a given sampling window, we 
have $L^{Rcv} = 1- N_i^{Rcv}/N_t^{RtrEg}$. 

Closely related to the previous one, we
also consider the \textbf{accuracy of received rate} by computing it based
on the \texttt{ip\_len} field of the reported packets, $R^{Rcv} =
\sum_0^{N} \mathtt{ip\_len}$.

Our last measured
variable is the \textbf{accuracy of timestamp reports} ($t_{Diff}^{Rcv}$). For a
given packet this is the difference between the timestamp from the \tcpdump\ 
traces and the one from the libtrace report at the receiver, $t_{Diff}^{Rcv} =
t_t^{Rcv} - t_i^{Rcv}$.

For the first and second set of experiments, we would like to stress that we are 
only interested
in the relative differences in our dependent variables in response to our
independent factors (\eg, OML instrumentation), rather than their absolute
``true'' values.  Indeed, running many measurement tools on a machine (\eg,
libtrace and \tcpdump\ on the receiver) will probably introduce biases in the
absolute measured value of performance or accuracy. However, we compare the
differences between runs where only our above factors vary and all others
parameters remain identical. Thus these biases have the same occurrence
probability across runs, and deriving claims on the analysis of the variances
from these runs is scientifically sound.\footnote{From a statistical point of
view, the variances induced by these biases will be part of the residuals in our
subsequent analysis.}

\subsubsection{Experiment Set 3}

This set aims at quantifying the effect of the OML measurement traffic 
on the experiment-generated traffic, when they both share the same wireless 
channel. In this case, an impact is indeed expected as both types of traffic 
contend for the same medium. In this set of experiments, we are interested in 
characterising how the experiment traffic varies as a function of the
amount of collected measurement.

The first factor in this set is the \textbf{OML sampling rate}, which is the
frequency (per sample) at which an OML-instrumented libtrace (\omltrace)
generates an aggregated report and sends it to the server. We vary
this factor in the range $\mathsf1$, $\mathsf{\frac{1}{10}}$,
$\mathsf{\frac{1}{50}}$, and $\mathsf{\frac{1}{100}}$---the higher the
sampling rate, the higher the number of measurements forwarded to the
server. The second factor considered is the \textbf{packet size} that Iperf uses
as the MTU for the experimental wireless path. Two values are considered,
\treatment{1,500} and \treatment{1,000}\,B, in order to explore the impact of a
varying number of packets on the same theoretically achievable throughput.

We measure a single dependent variable, which is the \textbf{achieved
throughput} as reported by an OML-instrumented Iperf in legacy mode on the
receiver.

\subsection{Experiment Execution}

In the first two sets, all experiments were performed on a testbed where the
nodes are all recent machines.\footnote{3.20\,GHz Pentium 4 processors, with
2\,GB of RAM running Ubuntu Linux with kernel \texttt{2.6.35-30-generic
\#59-Ubuntu}} Two separate Intel Pro/1000 Gigabit\footnote{It is important to
note that the PCI buses on our experimental machines was limited to 500\,Mbps
full-duplex. It was therefore not possible to achieve full Gigabit LAN traffic.}
LAN interfaces carried the experimental traffic on one side and the control and
measurement traffic on the other. This ensured that neither the control and
measurement traffic nor the performance capability of the machines biased our
experiments.  UDP was used to transport the experimental traffic to be able to
analyse jitters.

For the third set, we used a wireless testbed similar to
ORBIT~\cite{2005raychaudhuri_orbit}.  The wireless nodes were connected via an 802.11g adhoc
network, which carried both experimental and measurement traffic. TCP was used
to transport the experiment traffic as this is what OML streams use, thus
avoiding any potential TCP/UDP interaction bias.  The detailed specifications of
these nodes can be found in~\cite{2011mehani_oml_performance}. 

All our experiments were performed using the OMF
framework~\cite{2010rakotoarivelo_omf} on the IREEL experimentation
portal~\cite{2011jourjon_portal}.  Their precise descriptions are
available\footnote{\url{http://ireel.npc.nicta.com.au/projects/omlperf/wiki}}
and can be used to reproduce them on any OMF-enabled testbeds.

%% file: oml-datapath.pdftex_t
\begin{picture}(0,0)%
\includegraphics{oml-datapath.pdftex}%
\end{picture}%
\setlength{\unitlength}{3947sp}%
\begingroup\makeatletter\ifx\SetFigFont\undefined%
\gdef\SetFigFont#1#2#3#4#5{%
  \reset@font\fontsize{#1}{#2pt}%
  \fontfamily{#3}\fontseries{#4}\fontshape{#5}%
  \selectfont}%
\fi\endgroup%
\begin{picture}(7374,2799)(-11,-1948)
\put(5476,314){\makebox(0,0)[b]{\smash{{\SetFigFont{12}{14.4}{\sfdefault}{\mddefault}{\updefault}{\color[rgb]{0,0,0}OML}%
}}}}
\put(5476, 89){\makebox(0,0)[b]{\smash{{\SetFigFont{12}{14.4}{\sfdefault}{\mddefault}{\updefault}{\color[rgb]{0,0,0}Server}%
}}}}
\put(6901,-61){\makebox(0,0)[b]{\smash{{\SetFigFont{10}{12.0}{\sfdefault}{\mddefault}{\updefault}{\color[rgb]{0,0,0}database}%
}}}}
\put(6901, 89){\makebox(0,0)[b]{\smash{{\SetFigFont{10}{12.0}{\sfdefault}{\mddefault}{\updefault}{\color[rgb]{0,0,0}SQL}%
}}}}
\put(5476,-961){\makebox(0,0)[b]{\smash{{\SetFigFont{12}{14.4}{\sfdefault}{\mddefault}{\updefault}{\color[rgb]{0,0,0}OML}%
}}}}
\put(5476,-1186){\makebox(0,0)[b]{\smash{{\SetFigFont{12}{14.4}{\sfdefault}{\mddefault}{\updefault}{\color[rgb]{0,0,0}Server}%
}}}}
\put(6901,-1336){\makebox(0,0)[b]{\smash{{\SetFigFont{10}{12.0}{\sfdefault}{\mddefault}{\updefault}{\color[rgb]{0,0,0}database}%
}}}}
\put(6901,-1186){\makebox(0,0)[b]{\smash{{\SetFigFont{10}{12.0}{\sfdefault}{\mddefault}{\updefault}{\color[rgb]{0,0,0}SQL}%
}}}}
\put(4276,-1576){\makebox(0,0)[b]{\smash{{\SetFigFont{12}{14.4}{\sfdefault}{\mddefault}{\updefault}{\color[rgb]{0,0,0}file}%
}}}}
\put(4276,-1336){\makebox(0,0)[b]{\smash{{\SetFigFont{12}{14.4}{\sfdefault}{\mddefault}{\updefault}{\color[rgb]{0,0,0}Local}%
}}}}
\put( 76,614){\makebox(0,0)[lb]{\smash{{\SetFigFont{12}{14.4}{\sfdefault}{\mddefault}{\updefault}{\color[rgb]{0,0,0}App}%
}}}}
\put(826,-1336){\makebox(0,0)[b]{\smash{{\SetFigFont{12}{14.4}{\sfdefault}{\mddefault}{\updefault}{\color[rgb]{0,0,0}MS3}%
}}}}
\put(826,-586){\makebox(0,0)[b]{\smash{{\SetFigFont{12}{14.4}{\sfdefault}{\mddefault}{\updefault}{\color[rgb]{0,0,0}MS2}%
}}}}
\put(826,464){\makebox(0,0)[b]{\smash{{\SetFigFont{12}{14.4}{\sfdefault}{\mddefault}{\updefault}{\color[rgb]{0,0,0}MS1}%
}}}}
\put(1801,539){\makebox(0,0)[b]{\smash{{\SetFigFont{10}{12.0}{\sfdefault}{\mddefault}{\updefault}{\color[rgb]{0,0,0}Filters}%
}}}}
\put(3601,614){\makebox(0,0)[rb]{\smash{{\SetFigFont{12}{14.4}{\sfdefault}{\mddefault}{\updefault}{\color[rgb]{0,0,0}OML client library}%
}}}}
\put(601,-1636){\makebox(0,0)[b]{\smash{{\SetFigFont{12}{14.4}{\sfdefault}{\mddefault}{\updefault}{\color[rgb]{0,0,0}MP3}%
}}}}
\put(601,-886){\makebox(0,0)[b]{\smash{{\SetFigFont{12}{14.4}{\sfdefault}{\mddefault}{\updefault}{\color[rgb]{0,0,0}MP2}%
}}}}
\put(601, 14){\makebox(0,0)[b]{\smash{{\SetFigFont{12}{14.4}{\sfdefault}{\mddefault}{\updefault}{\color[rgb]{0,0,0}MP1}%
}}}}
\end{picture}%

%% file: 2012mehani_oml_performance_results.tex
\section{Results} 
\label{sec:results}

This section presents the results of the analyses which we performed on the
collected experimental data. All of the measurement data collected from our 
sets of experiments are available online with the R scripts used to analyse 
them.\footnotemark[\value{footnote}]

In each 5\,mn run, we collected 600 aggregate samples (one every half-second)
for each specific parameter cases in each experiment set, to mitigate any
unforeseen random factor. The sample times where rebased to 0 in their
respective timeframe to allow for meaningful comparisons between sources. The
first and last samples of each data sets were ignored to avoid bias due to
incomplete measurement periods at startup and teardown.

%

\subsection{Analysis and Assumptions}

An Analysis of Variance (ANOVA) is the established analysis for 
comparative studies where the factors have categorical levels and the dependent
variables are continuous. This is our case, as described in Section~\ref{sec:experiments}. Thus, we performed a series of ANOVAs on our collected data 
and attempt to disprove our null hypotheses by identifying significant 
variations between each cases in our experimental sets. We set our significance 
level $\alpha=0.05$ to have 95\,\% confidence when finding significant 
differences.\footnote{If a factor is found to have a significant impact, the 
probability of it being a false positive (Type I error) is < 5\%.} We note that 
some of the ANOVA assumptions are not always met by our data, and address 
them as follows.

\subsubsection{Independence of samples} As the samples of each variables for one trial
come from a time series, they are clearly not independent, which is confirmed by
Turning points tests~\cite{1989morley_statistical_tests_time-series}. We
therefore make our data iid by sampling it randomly with
replacement as suggested in~\cite{2010leboudec_performance}.\\

\subsubsection{Homoskedasticity} In some cases, Breusch-Pagan tests show that 
the variance of our samples differs significantly between treatment groups. 
Studies have however showed that the ANOVA is robust to deviations from this 
assumption at the price of a small reduction of the confidence $1-\alpha$ and an 
increase of the power of the test
$\beta$~\cite{1972glass_failure_meeting_anova_assumptions,1992harwell_monte-carlo_results_anova}.
Moreover, we note that these studies focused on ratio of variances only as low as
1:2. Even in our extreme cases, computing the ratio of the variances reveals
that the heteroskedasticity is much more modest than the cases studied in~\cite{1972glass_failure_meeting_anova_assumptions,1992harwell_monte-carlo_results_anova}.
We therefore conclude that our performed ANOVAs gave us valid results even with
this caveat on the confidence.\\

\subsubsection{Normality} This is the assumption from which our data
deviated the most, both in terms of skewness and kurtosis. We characterised 
this deviation with a Shapiro-Wilk test for each treatment group and proceeded
with an ANOVA if the deviation was not found to be significant. In case of
significant deviations, we used a non-parametric version of the ANOVA which 
removes the normality assumption by creating an empirical null distribution through permutation of the samples throughout treatments~\cite{2001anderson_permanova}.\\

\subsection{Experiment Set 1 (Iperf Instrumentation)}
\label{sec:results:iperf}

We performed two-way ANOVAs with interactions for each of the variables
$R_t^{RtrIn}$, $T_\mathrm{Diff}^{Rcv}$ and $J_\mathrm{Diff}$ at each of the
studied set rates. Characteristic results are presented below.

\subsubsection{Actual Sending Rate}

The results of the ANOVA for Iperf's sending rate as measured by an
un-instrumented \tcpdump, $R_t^{RtrIn}$, at rates \ftn, \fh\  and \ftth\,Mbps are shown in
Table~\ref{tab:iperfrate-anova} (left). The results of the analysis for set rate
\fft\ are similar to those for \ftn, while those for \fsf, and \fth\,Mbps are
similar to those for \fh.

\begin{landscape}

\begin{table*}[tb!]
  \addtolength{\tabcolsep}{-.25em}
  \centering
  \caption{Two-way PERMANOVAs with interactions on the actual sending rate
  of Iperf, $R_t^{RtrIn}$ (left), and the difference
  $T_\mathrm{Diff}^{Rcv}$ between the actual received rate and Iperf's throughput
  report (right).}
  \label{tab:iperfrate-anova}
  \begin{tabular}{lr|rrrrc|rrrrc}
    \toprule
    & & \multicolumn{5}{c|}{$R_t^{RtrIn}$} & \multicolumn{5}{c}{$T_\mathrm{Diff}^{Rcv}$} \\
    \cmidrule{1-12}
    & \textbf{d.f.} & \multicolumn{1}{c}{$SS$} & \multicolumn{1}{c}{$MS$} & \multicolumn{1}{c}{$F$} & \multicolumn{1}{c}{$p$} & \textbf{Signif.} &
       \multicolumn{1}{c}{$SS$} & \multicolumn{1}{c}{$MS$} & \multicolumn{1}{c}{$F$} & \multicolumn{1}{c}{$p$} & \textbf{Signif.} \\
    \midrule
    & \multicolumn{11}{c}{\textbf{\ftn\,Mbps}} \\
    \cmidrule{1-12}
    \textbf{oml} & $3$ & $2.69\times10^{6}$ & $8.97\times10^{5}$ & $1.76$ & $0.17$ & -- & 
      $1.12\times10^{8}$ & $3.72\times10^{7}$ & $0.95$ & $0.42$ & -- \\ 
    \textbf{threads} & $1$ & $9.86\times10^{5}$ & $9.86\times10^{5}$ & $1.94$ & $0.17$ & -- & 
      $1.19\times10^{7}$ & $1.19\times10^{7}$ & $0.30$ & $0.62$ & -- \\ 
    \textbf{oml:threads} & $3$ & $2.07\times10^{6}$ & $6.89\times10^{5}$ & $1.35$ & $0.26$ & -- & 
      $4.13\times10^{7}$ & $1.38\times10^{7}$ & $0.35$ & $0.83$ & -- \\ 
    \midrule
    & \multicolumn{11}{c}{\textbf{\fh\,Mbps}} \\
    \cmidrule{1-12}
    \textbf{oml} & 3 & $6.15\times10^{10}$ & $2.05\times10^{10}$ & $0.81$ & 1.00 & -- &
      $8.81\times10^{8}$ & $2.94\times10^{8}$ & 0.12 & $1.00$ & -- \\ 
    \textbf{threads} & 1 & $4.68\times10^{10}$ & $4.68\times10^{10}$ & 1.85 & 0.001 & $\star\star\star$ &
      $7.21\times10^{8}$ & $7.21\times10^{8}$ & 0.29 & $0.85$ & --\\ 
    \textbf{oml:threads} & 3 & $9.85\times10^{10}$ & $3.28\times10^{10}$ & 1.30 & 0.001 & $\star\star\star$ &
      $2.69\times10^{9}$ & $8.96\times10^{8}$ & 0.36 & $0.65$ & --\\ 
    \midrule
    & \multicolumn{11}{c}{\textbf{\ftth\,Mbps}} \\
    \cmidrule{1-12}
    \textbf{oml} & $3$ & $4.00\times10^{16}$ & $1.33\times10^{16}$ & $2.45\times10^{4}$ & 0.001 & $\star\star\star$ &
    $1.65\times10^{15}$ & $5.50\times10^{14}$ & $3.79\times10^{3}$ & 0.001 & $\star\star\star$ \\ 
    \textbf{threads} & $1$ & $8.76\times10^{14}$ & $8.76\times10^{14}$ & $1.61\times10^{3}$ & 0.001 & $\star\star\star$ & 
    $1.69\times10^{14}$ & $1.69\times10^{14}$ & $1.16\times10^{3}$ & 0.001 & $\star\star\star$ \\ 
    \textbf{oml:threads} & $3$ & $2.60\times10^{15}$ & $8.65\times10^{14}$ & $1.59\times10^{3}$ & 0.001 & $\star\star\star$ &
    $4.78\times10^{14}$ & $1.59\times10^{14}$ & $1.10\times10^{3}$ & 0.001 & $\star\star\star$ \\ 
    \bottomrule
    \addlinespace
    \multicolumn{7}{l}{Significance level: $\star\ 0.05$, $\star\star\ 0.01$, $\star\star\star\ 0.001$}
  \end{tabular}
\end{table*}
\end{landscape}

\begin{figure}[tbp]
  \centering
  \includegraphics[width=.75\columnwidth]{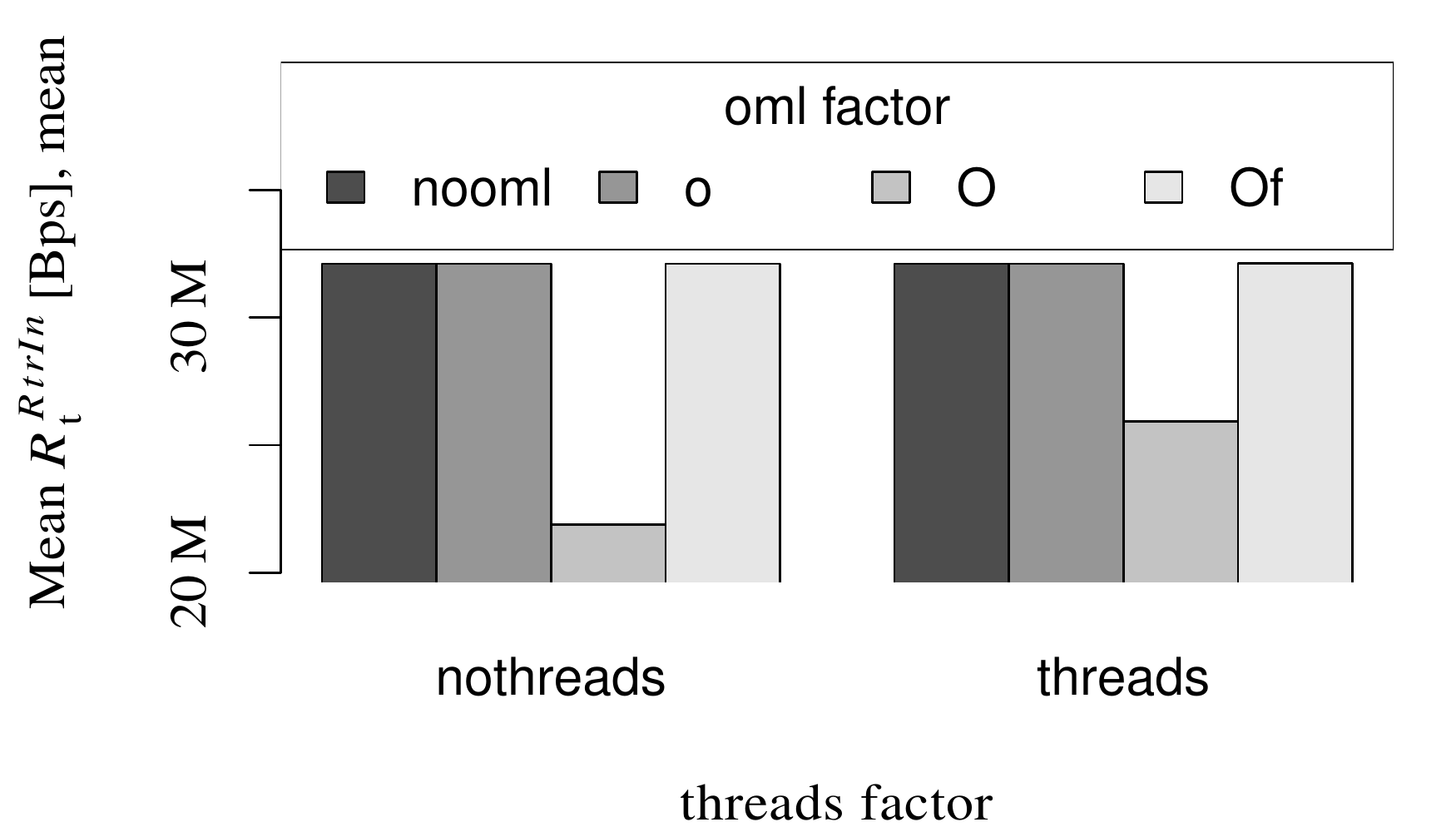}
  \caption{
  Sender rate $R_t^{RtrIn}$ achieved for different combinations of the
  experimental factors on
  %
  %
  at \ftth\,Mbps. While advanced OML appears to have a negative
  impact, it is partially mitigated by Iperf's internal threading, and fully
  removed by the use of OML filters (see Table~\ref{tab:dumpsnd-tukey}).}
  \label{fig:dumpsnd-interactions}

\end{figure}

For rates \fsf\,Mbps and higher, there are statistically significant differences
in Iperf's sending rates, which are introduced by changes in both the use of
threads and OML instrumentation, as well as the interaction of those two
factors. When significant, this interaction has to be studied first, which we do
in \figurename~\ref{fig:dumpsnd-interactions} for case \ftth\,Mbps. This figure
shows
%
%
the means of the response variable%
%
%
. It shows an
interaction between the threads and oml factors. The \Ooml\ factor introduces a
sizable negative impact, which is only partially mitigated by Iperf's native
threads. However, the use of filters in the \Ofoml\ factor completely removes
the issue.

The Tukey Honest Significant Differences test allows us to quantify the
deviations observed in \figurename~\ref{fig:dumpsnd-interactions}. We present
the relevant results allowing to characterise the previous figure in
Table~\ref{tab:dumpsnd-tukey}. For legibility's sake, we only show the mean
differences and the $p$-values. We however include all the differences between
interactions which were found to be significant.

\begin{table}[tb!]
  \addtolength{\tabcolsep}{-.3em}
  \centering
  \caption{Tukey Honest Significant Differences for oml:threads interactions
  with significant differences in the sending rate at \ftth\,Mbps. All 
  differences found significant ($p\le0.05$) are shown.}
  \label{tab:dumpsnd-tukey}
  \begin{tabular}{rrrc}
    \toprule
    \multicolumn{1}{c}{\textbf{oml:threads interaction}}& \multicolumn{1}{c}{\textbf{diff}}
    		& \multicolumn{1}{c}{\textbf{$p$ adj}} & \textbf{Signif.} \\ 
    & \multicolumn{1}{c}{\textbf{[MBps]}} \\
    \midrule
    \textbf{\Ooml:\nothreads--\nooml:\nothreads} & -$10.2$ & $0.00$ & $\star\star\star$ \\ 
    \textbf{\Ooml:\threads--\nooml:\nothreads} & -$6.08$ & $0.00$ & $\star\star\star$ \\ 
    \textbf{\Ooml:\nothreads--\Ofoml:\nothreads} & -$10.2$ & $0.00$ & $\star\star\star$ \\ 
    \textbf{\Ooml:\threads--\Ofoml:\nothreads} & -$6.08$ & $0.00$ & $\star\star\star$ \\ 
    \textbf{\Ooml:\nothreads--\ooml:\nothreads} & -$10.2$ & $0.00$ & $\star\star\star$ \\ 
    \textbf{\Ooml:\threads--\ooml:\nothreads} & -$6.08$ & $0.00$ & $\star\star\star$ \\ 
    \textbf{\nooml:\threads--\Ooml:\nothreads} & $10.2$ & $0.00$ & $\star\star\star$ \\ 
    \textbf{\Ofoml:\threads--\Ooml:\nothreads} & $10.3$ & $0.00$ & $\star\star\star$ \\ 
    \textbf{\ooml:\threads--\Ooml:\nothreads} & $10.2$ & $0.00$ & $\star\star\star$ \\ 
    \textbf{\Ooml:\threads--\Ooml:\nothreads} & $4.17$ & $0.00$ & $\star\star\star$ \\ 
    \textbf{\Ooml:\threads--\nooml:\threads} & -$6.08$ & $0.00$ & $\star\star\star$ \\ 
    \textbf{\Ooml:\threads--\Ofoml:\threads} & -$6.09$ & $0.00$ & $\star\star\star$ \\ 
    \textbf{\Ooml:\threads--\ooml:\threads} & -$6.08$ & $0.00$ & $\star\star\star$ \\ 
    \bottomrule
    \addlinespace
    \multicolumn{4}{l}{Significance level: $\star\ 0.05$, $\star\star\ 0.01$, $\star\star\star\ 0.001$}
  \end{tabular}
\end{table}

Table~\ref{tab:dumpsnd-tukey} confirms the observations from
\figurename~\ref{fig:dumpsnd-interactions}. Using OML to report every packets
(\Ooml) has a significant impact on the sending rate, reducing it by an average
(at most) 10.3\,MBps at set rate \ftth\,Mbps (about 27.5\%). Iperf's
internal threads can reduce this difference by 4.17\,MBps, which is still a
16.3\% drop from the performance for the other treatments.  Interestingly, though it is not found significant here, the use
of OML filters in the \Ofoml\ treatments consistently enabled a slight increase
in the sender's rate as compared to vanilla Iperf (\nooml\ treatment).

Next, we follow up with similar analyses for the other dependent variables we are
considering. However, due to space constraints, we report subsequent
results mostly in the text.

\subsubsection{Accuracy of Throughput Reports}

Here, we assess the variations of $T_\mathrm{Diff}^{Rcv}$ between the treatment groups,
as an evaluation of the impact of the instrumentation on the accuracy of Iperf's report. 
Table~\ref{tab:iperfrate-anova} (right) presents the corresponding ANOVA results.

At rates \ftn--\fth\,Mbps, no statistically significant ($p>0.05$) difference can
be found. Only for set rate \ftth\,Mbps do important ($p\le0.05$) deviations in the
mean appear. The combination of the oml and threads factors is, once again,
studied first. The 
interaction between factors
is qualitatively similar to \figurename~\ref{fig:dumpsnd-interactions}. Tukey
HSD tests confirm that the treatment causing this deviation is also the
non-filtered advanced mode (\Ooml) in both \threads\ and \nothreads\ treatments.
No other difference in mean between other treatments (particularly \nooml\ and
\Ofoml) is found to be significant.

\subsubsection{Accuracy of Jitter Reports}

We finally attempt to find differences in $J_\mathrm{Diff}$ in a similar fashion.
As OML does not currently provide a jitter-computing filter, we could not
consider treatment \Ofoml\ in this case. For treatment \Ooml, we post-processed
the packet records based on their arrival times to compute~\eqref{eq:jitter}.

For rates \ftn\ and \fft\,Mbps, no significant difference in the means could be
found (p>0.05). For rates \fsf\,Mbps and higher, however, the analyses of variance
identified statistically significant ($p\le0.05$) deviations. They were always
linked to the Iperf advanced mode (\Ooml), in comparison to the vanilla and
legacy report modes.

This difference can be explained in a similar fashion as for the throughput
reports where, with an increasing number of packets not being reported, the
computed metric loses accuracy.  A jitter-computing filter for OML would address
this issue in the same way as the sum filter did in the previous section.

\subsection{Experiment Set 2 (libtrace Instrumentation)}
\label{sec:results:libtrace}

We performed a similar analysis on the relevant dependant variables of our
packet-capture experiment. Characteristic results are reported thereafter.

\subsubsection{Accuracy of Packet Reports}

In this experiment, we are first interested in the loss ratio $L^{Rcv}$. In all
treatments of the rate factor (\fth--\ftth\,Mbps), the OML-instrumented
packet-capture application's reports deviate in a statistically significant
manner ($p\le0.001$) from the non instrumented version. We recall that this
application reports two samples (\texttt{ip} and \texttt{udp}) per captured
packet. With packets of size 1,498\,B, this induces a rate between 6,675 and
200,267\,pps, and double the number of samples. On average, 7.75\,pps went
unreported at \ftn\,Mbps (1.42\,\%), but this went up to 216\,pps at
\ftth\,Mbps (0.9\,\%). 

As no loss filter is currently available for OML, only the \nooml\ and \oml\
treatments were studied for the oml factor. Considering the reported rate
$R^{Rcv}$, as computed by summing the IP length of the reported factors allows
to provides some insight nonetheless. As for the losses, the throughput
exhibited significant differences ($p\le0.001$)  depending on whether it was
computed from \noomltrace\ or \omltrace's reports, with the latter being
consistently lower (between 0.09 and 2.21\,\%). However, complementing the use
of \omltrace\ with a summing filter (\Ofoml) produces statistically significant
($p\le0.001$ for treatments \fsf--\ftth\,Mbps)
\emph{positive} differences. We hypothesise that limiting processing in the main thread and
reducing the number of report packets to be sent allowed for more packets to be
read on time from the packet capture buffers, resulting in an increase in the reported throughput by 0.15--0.85\,\%
depending on the cases.  These observations are consistent with the Iperf
results from the previous section.

\subsubsection{Timestamp Accuracy and Precision}

The \noomltrace\ tool does not compute a local timestamp as OML does for
\omltrace. It is therefore not possible to obtain $t^{Rcv}_{Diff}$ in the
\nooml\ case for comparison. Rather, we only consider the \oml\ treatment.  and
give summary statistics for $t^{Rcv}_{Diff}$.  They are summarised in
Table~\ref{tab:trcvdiff}. For rate \fh\,Mbps and below, the time difference is
almost neglectable, with a maximum at 0.47\,$\mu$s, and a mean of about
100\,ns. It is interesting to note that some minimal differences are
\emph{negative}, which would hint that slightly different clocks are involved
in the kernel-land timestamping of PCAP packets and OML's userland
\texttt{gettimeofday(3)} requests.

\begin{table}
  \centering
  \begin{tabular}{crrrrr}
    \toprule
    \textbf{Set rate} &
    \multicolumn{5}{c}{\textbf{$t^{Rcv}_{Diff}$ [s]}} \\
    \cmidrule{2-6}
    \textbf{[Mbps]} & \multicolumn{1}{c}{\textbf{min.}} & \multicolumn{1}{c}{\textbf{med.}} & \multicolumn{1}{c}{\textbf{avg.}} & \multicolumn{1}{c}{\textbf{max.}} & \multicolumn{1}{c}{\textbf{sd}} \\
    \midrule
    \ftn  & -7.0\,n	& 0.072\,$\mu$	& 0.10\,$\mu$	& 0.47\,$\mu$	& 0.10\,$\mu$ \\
    \fft  & -5.5\,n	& 0.069\,$\mu$	& 0.098\,$\mu$	& 0.47\,$\mu$	& 0.099\,$\mu$ \\
    \fsf  & -5.7\,n	& 0.077\,$\mu$	& 0.11\,$\mu$	& 0.47\,$\mu$	& 0.11\,$\mu$ \\
    \fh   & -7.0\,n	& 0.071\,$\mu$	& 0.10\,$\mu$	& 0.47\,$\mu$	& 0.10\,$\mu$ \\
    \fth  & 0.0		& 0.0		& 3.7		& 132.9		& 21.9 \\
    \ftth & 0.0		& 0.0		& 6.3		& 162.8		& 31.4 \\
    \bottomrule
  \end{tabular}
  \caption{Packet timestamp difference between PCAP capture via libtrace and
  OML report timestamp (\texttt{oml\_ts\_client}) generated by \omltrace.}
  \label{tab:trcvdiff}
\end{table}

The picture is clearly different for rates \fth\ and \ftth\,Mbps. For these
treatments, the maximum difference is more than 2 minutes and the average time
differences are of the order of several seconds. We recall that at these rates,
250,000 to 400,000 samples are generated per second. These samples are stored,
on the server side, in a FIFO queue, before being entered in a database. We
hypothetise that this is an indication that the OML server cannot handle such
high loads and effectively breaks before that.

\subsection{Experiment Set 3 (Reporting Traffic)}
\label{sec:results:sharedmedium}

\begin{figure}[tbp]
  \centering
  \includegraphics[width=.75\columnwidth]{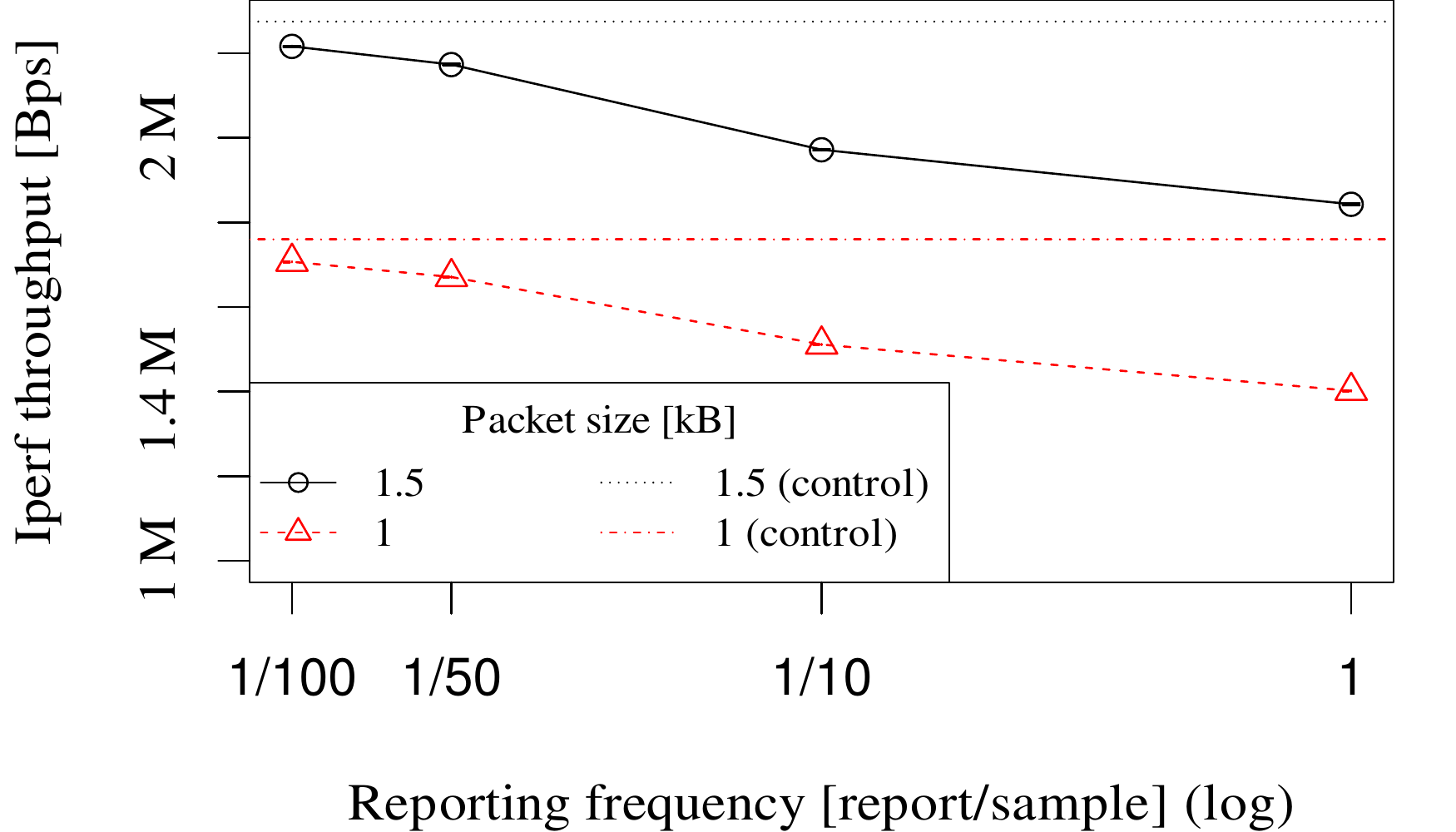}
  \caption{Mean achievable Iperf rate over 802.11g depending on the reporting
  rate of OML on the same medium (error bars show the standard deviation, but
  are barely visible with due to the $y$-axis scale in used).}
  \label{fig:rate-wifi}
\end{figure}

In this experiment set, we focused on Iperf's received TCP rate whilst OML was
used within \omltrace\ and configured to capture information from every  packets
and send reports on the same wireless network as the Iperf traffic, with varying
frequencies. Results are summarised in \figurename~\ref{fig:rate-wifi}. It shows
the mean measured throughputs (over ten runs), and their associated standard
deviation on the $y$-axis as a function of the OML reporting frequency, in
report per sample. In this scale, a frequency of $0.01$ indicates that the
measurement library only reports once every hundred captured packets. This still
represents many reports per seconds. Both treatments \treatment{1,500} and
\treatment{1,000}\,B of the packet sizes are shown. The behaviour at both sizes
is qualitatively similar. The figure also includes Iperf's upper-bounds
throughput in our wireless environment, measured with \omltrace\ using a control
network rather than the experimental radio network.

The results presented in \figurename~\ref{fig:rate-wifi} show that, as expected, 
the reporting traffic from the OML-instrumented applications impacts
the behaviour of the different experiments in a significant manner. This is consistent
with the behaviour of two flows sharing the same network. 
However, these results also demonstrate that, when OML filters are used---in this
case, to achieve 
differentiated sampling policies---these significant impacts over the 
experiments can be reduced.


%% file: 2012mehani_oml_performance_discussion.tex
\section{Discussion}
\label{sec:discussion}

In this section, we discuss some of the findings from the previous result 
analysis and offer some recommendations on the use of OML and the 
instrumentation and measurement collection in networking experiments.

\subsection{On OML's Observer Effect}

As mentioned earlier in Section~\ref{sec:method:objective}, our main objective
was to characterise the impact of OML on two types of tools widely used in the 
networking community, a network probing tool (Iperf) and a packet capture 
library (libtrace).

As a reminder, we defined in Section~\ref{sec:experiments} our study's two null
hypotheses as follows: ``the OML instrumentation of
Iperf has no significant effect on its packet-sending rate nor on the accuracy
of its throughput and jitter reports'' and ``the
OML instrumentation of libtrace has no significant effect on the accuracy of its
packet and timestamp reports.''

In the case of the Iperf's instrumentation, the results from Section
\ref{sec:results:iperf} show that the OML-instrumentation of the legacy
reporting mode of Iperf (treatment \ooml) does not introduce any significant
deviation from the normal behaviour, at any rate. However the advanced per
packet reporting mode (\Ooml) is a more intrusive and does introduce
a negative bias in the reported metrics and the general behaviour of the
application when used without care. The introduction in OML's reporting loop of aggregating functions such as a sum filter (\Ofoml) completely alleviates the
issue.

For the libtrace instrumentation, the results from
Section~\ref{sec:results:libtrace} show a similar trend, where per-packets
reports at very high rates have significant differences from what would be
measured without OML. Once again, the proper use of upstream processing filters
 can cancel out
this problem. Moreover in this case, we found a statistically significant
positive bias introduced by the use of filters, as it allowed the
packet-capturing tool to resume reading the capture buffer faster, while OML
was processing the samples in a separate thread.

In a similar manner, in the context of a wireless experimentation without
the availability of a separate communication channel to stream measurements, OML significantly
disturbs the experiments when it is configured to capture and report every
packet information. This behaviour was expected but our experiments show that
the upstream processing allows to asymptotically approach the performance
obtained when a dedicate reporting channel was available. These results 
demonstrate one of the
advantage of the OML filtering capabilities and we can envision that this
library would be a good candidate in the context of wireless
experimentations. 

Whilst the experiments presented in this study focused on the OML effect on the
instrumented measurement tools, it would be legitimate to also question the
effect of this library on the physical resources themselves. In order to
evaluate this effect, we performed a pilot
study~\cite{2011mehani_oml_performance} and found no significant impact on
neither the CPU nor the memory usage when both applications were reporting with
OML.  For the sake of space and clarity, we opted not to include these results
in the present document.

It is worth noting that, on the less powerful machines we used in that first
study, we also found a positive significant impact of OML when used with
thread-less versions of Iperf. In these cases, using OML instead of Iperf's
normal report channels would bring the performance of the non-threaded flavours
to that of the threaded ones. We direct the reader
to~\cite{2011mehani_oml_performance} for details.

Next, we propose a set of recommendations for the proper instrumentation and
control network. Nevertheless, we measure the pick traffic between the receiver and the 
usage of OML applications.

\subsection{Recommendations}

OML is a good candidate for the instrumentation of distributed wireless 
systems. Indeed, it unifies the collection of measurements from multiple 
distributed nodes, and simplifies the development of measurement application
(\eg, it removes the need for a complex threaded reporting system within the 
application). Our analyses suggest that significant impacts of OML 
can be avoided with adequate experimental and instrumentation design. We believe
that these considerations are not specific to OML and can be applied to 
a larger group of instrumentation frameworks, and we further discuss them below.

First, a developer should carefully select the
metrics to report together in a single MP when instrumenting either an 
existing or new piece of software.
It is a trade-off between providing flexibility to 
the future user (\ie, the experimenter) and limiting the volume of information
that may be reported to the collection
servers. For example, in some cases it may be relevant to report all possible 
metrics for events always arriving at the same time (\eg, metrics about every incoming 
packets), whereas in other cases only some aggregate metrics may be of 
interests (\eg, throughput or jitter).

This point highlights the decoupling of instrumentation concerns from 
measurement ones, which we believe is a desirable feature of an instrumentation 
framework. 
This allows the person
in charge of instrumenting an application to 
expose as many measurement points as possible without any assumptions on the experiments which 
will use it.
Thus, the experimenters retain the latitude of selecting
only the relevant MPs for their study, without having to tinker with the
application's code any further. Only the selected MPs 
would then generate measurements.
Such decoupling enables the reuse of instrumented applications in 
various types of studies, and gives the experimenters the final choice
on what data to collect. As a simple example, a researcher working on wireless 
MAC protocol could enable only the \verb_radiotap_ MP when using our OML-instrumented \omltrace\, while another researcher working on an application 
protocol and using the same tool could enable only the \verb_udp_ MP. 

An experimenter would also have to ensure that the volume
of measurements to collect is within the capacity of the collection server.
Several solutions exist to do so, such as enabling only the
relevant MPs or choosing an adequate sampling rate. In
this study we analysed another solution based on the use of OML
filters to pre-process data on the client side and thus reduce the
number of samples. While our previous results only showed the benefits of using a sum filter over a 
period of time to control the volume of collected data, we believe that other 
kinds of filters could have similar effects.  Using filters however assumes that 
the 
experimenters have an \apriori\
idea of which metrics may be relevant to their study. It might therefore not be
suitable in the exploratory phase of a study. In this case, 
selectively distributing the MSs between multiple collection servers would help 
control the load on the collection infrastructure. 

Another point for the experimenter to consider is the cost of sampling.
Indeed, while computing, storage, and network resources are often
considered inexpensive, collecting all available raw measurements in anything
than a simple experiments very often have a real cost in future data analysis
and management. For example, a lot of a researcher's time may be spent in
sorting or selecting relevant data from a large measurement set, while 
selective sampling, perhaps motivated by a prior small-scale pilot study, may have
produced more concise and relevant measurement sets. The use of filters allows
this by letting the
experimenters 
aggregate samples at different resolutions, thus giving them a fine control on
the trade-off between the amount and relevance of collected data.

%
%



%% file: 2012mehani_oml_performance_related.tex
\section{Related Work}
\label{sec:related}


Studies based on experimental measurements form a large part of the research in
networking. Many of these studies suffer from errors related to the measurement
tools or frameworks being
used~\cite{2004paxson_internet_measurement,2011krishnamurthy_socratic_measurement_validation}.  This may be
in part due to the fact that not all used instrumentation solutions have been
systematically evaluated for their impact on the system under study.  Even
though, examples of such thorough observer effect studies
exist~\cite{2005choi_observations_netflow,2010braun_packet_capture}.

Several solutions exist to instrument and collect information from 
networking applications and devices, such as SNMP~\cite{rfc3411} or
DTrace~\cite{2004cantrill_dtrace}. Similar to OML, they both allow the instrumentation of any 
software and/or devices. In addition, DTrace can dynamically instrument live 
applications, and is shipped by default with some operating systems. However, 
its measurement processing is limited to aggregating functions, and 
it does not support the streaming of measurements from different devices to a 
remote collection point. SNMP has been widely adopted for the management and 
monitoring of devices, and allows the collection of information over the 
network. However, it has some performance and scaling limitations when 
measurements from large number of devices are required within a short time 
window~\cite{2006zhao_traffic_estimation}.

IPFIX~\cite{rfc5101} is an IETF standard, which defines a protocol for 
streaming information about IP traffic over the network. Similar to OML clients, 
IPFIX exporters stream collected and potentially filtered measurements to 
collector points. However, IPFIX is limited to measurements about IP flows. OML 
currently provides the choice of two protocols to stream measurements, a text-
based and a binary one, and has the support for the IPFIX protocol on its 
development roadmap.\footnote{\url{http://oml.mytestbed.net/projects/oml/
roadmap}}

The networking community has been developing and using several measurement 
tools, from high performance or versatile devices such as 
DAG\footnote{\url{http://www.endace.com}} or
NetFPGA~\cite{2008gibb_netfpga_teaching} to
specialised or distributed software such as 
Radiotap\footnote{\url{http://www.radiotap.org}} or DIMES~\cite{2005shavitt_dimes}.
Most of these tools could be instrumented with OML, \ie, as a streaming and
collection framework for the data that they produce. This would allow the tools' 
users to benefit from features such as filtering close to the source, easy 
correlation of data from many sources through the timestamped collection at one 
or many points, or support for temporary disconnection (\eg, DIMES 
agents on laptops).

%% file: 2012mehani_oml_performance_conclusion.tex
\section{Conclusion} 
\label{sec:conclusion}

In this article, we characterised the observer effect induced by instrumentation
framework on measurement tools commonly used in networking research.
In that regard, we proposed a methodology, based on a few easy steps and the use
of
analysis of variance techniques, to quantify and compare the deviations between
the original tools and their instrumented counterparts. This methodology
can be applied to any measurement frameworks.

In our case, we applied this methodology to analyse the differences in
performance and accuracy of reported metrics.
Our results showed that, though some significant negative impacts due to the
OML-instrumentation could be found, proper setup of the collection system could
entirely prevent them.  Moreover, we identified some statistically significant
positive effects, when using the in-stream filtering capabilities of OML, in
some cases where the instrumented application did not use threads. Indeed, OML
removes the complexity of reporting from an application, thus facilitating its
development and providing users with a non-intrusive way to collect metrics from
applications which main purpose is not necessary measurement.  We also
identified some limitations in the current OML collection server.
 
Finally, we presented some recommendations for developers to instrument 
their software, and for experimenters to configure the measurement 
collection from these software to avoid impacting their performance.
